\documentclass{article}

\def\ignore#1{}

\def\psfancypar#1#2{\begingroup\def\par{\endgraf\endgroup\lineskiplimit=0pt}
               \setbox2=\hbox{\large\sc #2}
               \newdimen\tmpht \tmpht \ht2 \advance\tmpht by \baselineskip
               \font\hhuge=Times-Bold at \tmpht
               \setbox1=\hbox{{\hhuge #1}}
               \count7=\tmpht \count8=\ht1
               \divide\count8 by 1000 \divide\count7 by \count8 
               \tmpht=.001\tmpht\multiply\tmpht by \count7 
               \font\hhuge=Times-Bold at \tmpht
               \setbox1=\hbox{{\hhuge #1}}
               \noindent
                \hangindent1.05\wd1
               \hangafter=-2 {\hskip-\hangindent
               \lower1\ht1\hbox{\raise1.0\ht2\copy1}%
                \kern-0\wd1}\copy2\lineskiplimit=-1000pt}

\newcommand{\E}{\mbox{{\rm E}}}

\def\boxit#1{\vbox{\hrule\hbox{\vrule\kern3pt
        \vbox{\kern3pt#1\kern3pt}\kern3pt\vrule}\hrule}}

\def\reals{ { {\rm  I \kern-0.15em R }  } }
\def\complex{ {\,{{\rm C} \kern-0.50em \raise0.20ex {  |}}\, }}

\def\mubf{\hbox{\boldmath$\mu$\unboldmath}}

\def\Qbf{{\bf Q}}
\def\Rbf{{\bf R}}

\def\Xbf{{\bf X}}

\def\Ic{{\cal I}}

\def\be{\vskip .3cm \begin{equation}}
\def\ee{\end{equation} \vskip .4cm \noindent}
\def\defeq{{\stackrel{\Delta}{=}}}
\newcommand{\R}{\mbox{$\hat {\bf R}_{N}$}}

\def\Rxx{\Rbf_{\ssstyle X\kern-.1em X}}

\let\ssstyle=\scriptscriptstyle

\def\Kout{\setbox1=\hbox{\Huge\bf K}\hbox to
1.05\wd1{\hspace{.05\wd1}
}}
\def\Sout{\setbox1=\hbox{\Huge\bf S}\hbox to 1.05\wd1{\hspace{.05\wd1}
}}

\input setup
  \ifx\LabelFigloaded\MYundefined\relax
  \else
   \fi

  \def\LabelFigloaded{\relax}% now loaded

  \chardef\LabelFigCatAt\the\catcode`\@
  \catcode`\@=11

 \let\LabelFigwlog@ld\wlog
 \def\wlog#1{\relax}

 \ifx\\\MYundefined@
    \let\\\relax
 \fi

  \def\ms@g{\immediate\write16}

 \def\N@wif{\csname newif\endcsname }
 \def\Temp@ {\N@wif\ifIN@}
 \ifx\INN@\MYundefined@
    \else \let\Temp@\relax
 \fi
 \Temp@

  \def\IN@{\expandafter\INN@\expandafter}
  \long\def\INN@0#1@#2@{\long\def\NI@##1#1##2##3\ENDNI@
    {\ifx\m@rker##2\IN@false\else\IN@true\fi}%
     \expandafter\NI@#2@@#1\m@rker\ENDNI@}
  \def\m@rker{\m@@rker}
 
  \newtoks\Initialtoks@  \newtoks\Terminaltoks@
  \def\SPLIT@{\expandafter\SPLITT@\expandafter}
  \def\SPLITT@0#1@#2@{\def\TTILPS@##1#1##2@{%
     \Initialtoks@{##1}\Terminaltoks@{##2}}\expandafter\TTILPS@#2@}

 \def\Shifted@@#1#2#3{\setbox0=\hbox{#3}%
   \raise -\dp0\vbox {\kern-#2%
       \hbox {\kern#1\unhbox0\kern-#1}%
           \kern#2}}

 \newcount\gridcount
 \newbox\auxGridbox@ \newbox\hGridbox@ \newbox\vGridbox@
 \newbox\Labelbox@ \newbox\auxLabelbox@
 \newbox\Coordinatebox@
 \newtoks\Labeltoks@
 \newdimen\Wdd@ \newdimen\Htt@
 \newdimen\Wddd@ \newdimen\Httt@
 
 \def\Wr@{\immediate\write16}

 \newdimen\GL@wd%% grid-line width
 \def\GridLineWidth#1{\GL@wd=#1}

 \def\gobble#1{}
 \def\EdgeErr@{\Wr@{}%
      \Wr@{\string\Edges\space argument
      1, 10, 100 or 1000 please\string!}%
      }

 \newcount\Edgect@

 \def\Sweepup#1\endSweepup{}

 \def\SetEdges@{%
    \edef\Zr@@s{\expandafter\gobble\number\Edgect@\empty}%
        \count255=0\Zr@@s\relax
        \ifnum\count255=\z@\else\EdgeErr@\show\tailtest\fi
        \count255=1\Zr@@s\relax%\showthe\count255
        \ifnum\count255=\Edgect@\relax\else\EdgeErr@\show\leadtest\fi
    \EdgGl@b\edef\Zr@s{\expandafter\gobble\Zr@@s\empty}%\show\Zr@s
    \ifnum\Edgect@>\@ne\relax\EdgGl@b\let\L@Dc\empty
        \else\EdgGl@b\edef\L@Dc{\string.}\fi
    \ifnum\Edgect@>\@ne\relax
        \EdgGl@b\edef\Edgescale@##1{\divide##1 by \Edgect@}%
        \else\EdgGl@b\edef\Edgescale@##1{}\fi
    }

 \def\Edges#1{\Edgect@=#1\relax
     \let\EdgGl@b\global \SetEdges@}

 \Edges{1}%% default

 \def\hhrule{\hrule height \GL@wd\vskip-.\GL@wd}

 \def\hRule@{%
   \advance\gridcount -2%
   \vfil\hhrule\vfil
   \llap{\smash{\raise -2.5pt
     \hbox{\L@Dc\number\gridcount\Zr@s\kern2pt}}}%
   \hhrule
   }

\def\vvrule{\vrule width \GL@wd \kern-\GL@wd}

 \def\vRule@{\advance\gridcount 2%
   \hfil\vvrule\hfil
   \setbox\auxGridbox@=\vbox to 0pt
      {\vskip \Htt@\vskip 2pt
        \hbox to 0pt{\hss\L@Dc\number\gridcount\Zr@s\hss}\vss}%
      \wd\auxGridbox@=0pt \box\auxGridbox@
   \vvrule
   }

 \def\PlaceGrid@@{\gridcount=10 
  \setbox\hGridbox@=\hbox{%
        \hbox{%
             \hskip-.4pt\vrule
             \vbox to \Htt@{%
               \offinterlineskip\parindent=\z@\relax
               \hbox to \Wdd@{\hfil}
               \hRule@\hRule@\hRule@\hRule@
               \vfil\hhrule\vfil}%
             \vrule\hskip-.4pt}
    }%
  \gridcount=0%
  \setbox\vGridbox@=\hbox{%
      \vbox{\offinterlineskip\parindent=0pt\hsize=0pt
         \vskip-.4pt\hrule%
         \hbox to \Wdd@{%
                 \vtop to \Htt@{\vfil}%
                 \vRule@\vRule@\vRule@\vRule@
                 \hfil\vvrule\hfil}%
         \hrule\vskip-.4pt}}%
  \wd\hGridbox@=0pt\ht\hGridbox@=0pt
  \wd\vGridbox@=0pt\ht\vGridbox@=0pt
  \hbox{\box\hGridbox@\box\vGridbox@}%
  }

 \def\LabelsGlobal{\def\LabGl@b{\global}}
 \def\LabelsLocal{\def\LabGl@b{}}
 \LabelsGlobal %% default

 \def\SetLabels#1\endSetLabels{%
   \LabGl@b\Labeltoks@={#1()\\}%
   }

 \LabGl@b\Labeltoks@={()\\}

 \def\ShowGrid{\LabGl@b\let\PlaceGrid@\PlaceGrid@@}
 \def\HideGrid{\LabGl@b\let\PlaceGrid@\relax}
 \def\Grids{\ShowGrid\LabGl@b\let\GridSwitch@\ShowGrid}
 \def\noGrids{\HideGrid\LabGl@b\let\GridSwitch@\HideGrid}

 \noGrids

 \def\bAdjust@@{%
     \setbox\auxLabelbox@=\hbox{\raise \dp\auxLabelbox@
            \box\auxLabelbox@}}
 \def\bAdjust@{\let\vAdjust@\bAdjust@@}

 \def\eAdjust@@{\dimen0=-.5\ht\auxLabelbox@
     \advance\dimen0 by .5\dp\auxLabelbox@
     \setbox\auxLabelbox@=
            \hbox{\raise\dimen0\box\auxLabelbox@}}
 \def\eAdjust@{\let\vAdjust@\eAdjust@@}

 \def\tAdjust@@{%
     \setbox\auxLabelbox@=\hbox{\raise-\ht\auxLabelbox@
            \box\auxLabelbox@}}
 \def\tAdjust@{\let\vAdjust@\tAdjust@@}

 \let\vAdjust@\relax

 \def\lAdjust@{\let\hAdjust@\rlap}
 \def\rAdjust@{\let\hAdjust@\llap}

 \let\hAdjust@\relax\let\vAdjust@\relax

 \def\FetchLabel@#1(#2)#3\\{%
     \IN@0#2@@\ifIN@
        \setbox0=\hbox{\ignorespaces#1#3\unskip}%
        \ifdim\wd0>0pt
           \ms@g{}%
           \ms@g{ !!! Bad label(s)? !!!}%
           \message{ #1(#2)#3}%
        \fi
        \def\LabelMole@##1\endFetchLabel@{%
            \IN@0()\\@##1@%
            \ifIN@\def\Temp@{\FetchLabel@##1\endFetchLabel@}%
            \else\def\Temp@{}%
            \fi
            \Temp@
           }%
     \else
       \ignorespaces#1\unskip
       \setbox\auxLabelbox@=%
         \hbox to 0pt{\hss\ignorespaces\hAdjust@
          {\ignorespaces#3\unskip}\hss}%
       \vAdjust@
       \let\hAdjust@\relax\let\vAdjust@\relax
       \AugmentLabelBox@@{#2}%
       \ht\Labelbox@=0pt\dp\Labelbox@=0pt
       \let\LabelMole@\FetchLabel@%
     \fi\LabelMole@}

 \newtoks\XYSep@ %\XYSep@{*}
 \def\SetXYSeparator#1{%
     \IN@0#1@@\ifIN@\XYSep@{*}%
     \else
     \XYSep@{#1}%
     \fi
     }

 \SetXYSeparator*

 \def\AugmentLabelBox@@#1{%
     \IN@0\the\XYSep@ @#1@\ifIN@
       \SPLIT@0\the\XYSep@ @#1@%
       \setbox\Labelbox@=\hbox to 0pt{%
         \unhbox\Labelbox@
         \Shifted@@{\the\Initialtoks@\Wddd@}%
         {\the\Terminaltoks@\Httt@}%
         {\box\auxLabelbox@}}%
     \else
         \ms@g{}%
         \ms@g{ !!! Bad insertion point. !!!}%
         \message{ (#1\ this point was rejected.)}%
     \fi
    }

 \def\FetchOption@#1[#2]#3\endFetchOption@{%
    \def\temp{#1}%\show\temp
    \ifx\temp\empty
       \Edgect@=#2\relax%\showthe\Edgect@
       \let\EdgGl@b\relax
       \SetEdges@%\def\Edgescale@##1{\divide##1 by \Edgect@\relax}%
       \Cleaner@#3%
    \fi}

 \def\Cleaner@#1[@]{\Labeltoks@{#1}}
     
 \def\PlaceLabels@@{\mathsurround=0pt%\bgroup
     \def\Cr@{\\}%
     \let\L\lAdjust@\let\R\rAdjust@
     \let\B\bAdjust@\let\E\eAdjust@\let\T\tAdjust@
     \expandafter\FetchOption@\the\Labeltoks@[@]\endFetchOption@
     \Wddd@=\Wdd@ \Edgescale@\Wddd@ %\showthe\Edgect@
     \Httt@=\Htt@ \Edgescale@\Httt@
     \expandafter\FetchLabel@\the\Labeltoks@\endFetchLabel@
     \box\Labelbox@%\egroup
     }%

 \let \PlaceLabels@\PlaceLabels@@

 \def\AffixLabels#1{\setbox\Coordinatebox@=\hbox{#1}%
      \Wdd@=\wd\Coordinatebox@ \Htt@=\ht\Coordinatebox@
      \advance\Htt@ \dp\Coordinatebox@
      \hbox{\copy\Coordinatebox@\kern-\Wdd@ 
           \Shifted@@{0pt}{-\dp\Coordinatebox@}%
           {\PlaceLabels@\PlaceGrid@}%
           \kern\Wdd@}%
      \GridSwitch@ %% next grid hidden
      \LabGl@b\Labeltoks@{()\\}%
      }
 
   \let\wlog\LabelFigwlog@ld   %%restore logging
   \catcode`\@=\LabelFigCatAt  %%12 or 13

                                By

              Raymond S\'eroul <A18645@FRCCSC21.BITNET>
                                and 
              Laurent Siebenmann <lcs@topo.math.u-psud.fr>
    
              VERSIONS: July 1991, Oct 1991, Jan 1992, July 1992

INTRODUCTION

      This labelling package is intended for TeX users who
rely on non-TeX sources for for their graphics inserts.  It
provides means for adding TeX labels to such inserts with a
minimum of fuss. 

       For most labels, TeX users have in the past found it
reasonably convenient to rely on non-TeX sources. Typical
occasions when an inescapable need for TeX labels seemed to
arise are

 (a) when the graphics program lacks certain exotic or complex
mathematical symbols

 (b) when the very highest typographical quality is wanted for the
labels

 (c) when labels included with the graphics fail to print, 
 and you cannot figure out why (cf. boxedeps.doc).  The labels
 provided by labelfig.tex are 100% portable.

       Since this package first appeared, many users, who in the
past scarcely dreamed of using TeX labels, have come to use
nothing but.  So it is now appropriate to add

Intoxication Warning:  TeX labels may be addictive and expensive. 

     If you have a fast preview you may disagree, and even find
that this package provides an agreeable paste-up environment; see
extra applications at end.

     Note to publishers: It is possible and convenient to ultimately
export the TeX labels produced by labelfig.tex to become an integral
part of the EPS file. This is often desired by a publisher who typically
uses an "upmarket" graphics or page layout program, with which the
staff is skilled in perfecting figures.  See Appendix I for
a recipe.

     The authors are grateful to Patrick Ion of Math Reviews for
helpful comments and encouragement.

BASIC INSTRUCTIONS

    After reading in the macro file using

preview or proof your figure with a coordinate grid printed on
top, by typing the following:

    \ShowGrid  % shows grid  for next figure only
    \AffixLabels{<the graphics insertion>}

Here <the graphics insertion> is what you would type to insert
the graphics object alone without the grid.  This must provide
for the space around it. For example <the graphics insertion>
might well be \BoxedEPSF{MyFigure scaled 700} using the
boxedeps.tex macro package (from same source); this provides a
TeX box containing the encapsulated PostScript insert specified by
the file MyFigure. \AffixLabels{...} provides the grid (supposing
\ShowGrid is present) and later, once you have specified labels
using the grid, it will "tack on" the labels.

     The grid is a sort of (usually elongated) checkerboard of
ten rows and ten columns and its (internal) partitions are by
default numbered  .1, ... ,.9  both horizontally (X-coordinate
running left to right) and vertically (Y-coordinate running bottom
to top).  Thus the points enclosed by the grid correspond to the
points of the unit square in the cartesian "X-Y" plane, the lower
left corner corresponding to the origin (0,0).  By extrapolation,
the full page corresponds to a larger rectangle in the plane.

     These coordinates serve to position labels as follows.
Before the \AffixLabels{...} command type label specifications:

  \SetLabels
   (<X-coordinate>*<Y-coordinate>) <first label> \\
   .
   .
   .
   (<X-coordinate>*<Y-coordinate>)  <last label> \\
  \endSetLabels

Each row specifies one label and is terminated by \\.  In each
row, the position indicator comes first; it is written as a
standard cartesian point except that the X- and Y- coordinates
are separated by * rather than a comma because TeX allows a
comma as decimal point. There are no dimension units to specify
as the unit is the grid itself.

     By default, this cartesian point specifies where the middle
of the baseline of the label will be located.  However if you precede
the point by \L [or \R] the left [or right] edge of the baseline will
be located there. Similarly you may also precede the point by \T, \E,
or \B to vertically align the top equator or bottom of the label box
at the specified point.  This gives nine standard positions of
the label with respect to the insertion point --- corresponding to
the eight principle points of the compas and the center

                     \L\T     \T      \R\T

                     \L\E     \E      \R\E

                     \L\B     \B      \R\B

But this neglects the default "baseline" level of TeX,
giving potentially three more positions

                     \L    <no tag>   \R

For text, the baseline level is often the preferred. Its relation to
the others is variable. It will often coincide with the bottom level,
as happens for "X".  But it is often distinct, as for "g", in which
case you have in all 12 distinct positions rather than 9.

     It is convenient to think of this specification of label
position as attaching the label by a thumb-tack to the coordinate
grid. There are up to twelve positions of the thumb-tack on the
label, while the position of the thumb-tack on the coordinate grid is
arbitrary.  Normally, one choses the position of the thumb-tack on
the label to be the one that is the closest to the item being
labeled.  There are good reasons for this "rule of thumb":

   (a)  It facilitates correct positioning at first try.

   (b)  If the scale of the figure must be altered after labels
have been affixed, the labels have a good chance of remaining well
positioned.

   (c)  The visible grid need not extend beyond the "bounding box"
for the figure, because the best preferred position is always
(at least almost) within the bounding box .

The second reason is particularly important. Indeed it often
happens that scale has to be altered after labelling begins, in
order to either provide space for the labels, or to adjust
proportions between the labels and the figure.  (The size of labels
is unaffected by scaling.)

     Here is an artificial but self-contained test which uses
TeX rules to make a graphics object.

TEST

    Do not skip this!

 \def\FrameIt#1{\hbox{\vrule$\vcenter {\hrule\kern3pt%
             \hbox {\kern3pt #1\kern3pt}%
               \kern3pt\hrule}$\relax\vrule}}

 \def\Caption#1#2{\FrameIt{%
       \vtop {\hsize=#1\relax \parindent=0pt
         \leftskip=0pt \rightskip=0pt plus15pt
         \parfillskip=0pt
         \lineskip=1pt\baselineskip=0pt
         #2}}}

 \def\FirstQuadrant{\hbox to 100pt{\vrule\vbox to 100pt{%
        \hbox to 100pt{\hfil}\vfil\hrule}\hss}}

  \SetLabels
    \R(.5*.2) $\zeta\,\cdot$\\
    (.9*-.10) $\xi$\\
    \R(-.03*.9) $\eta$\\
    \T(.5*.9) \Caption{70pt}{%
          \it The norm of
          $g(\xi+i\eta)$ is indicated on
          contours of this invisible surface.}\\
  \endSetLabels

  \AffixLabels{\FirstQuadrant}

  \end

  Note that the coordinates to use for labels are indicated on the
edges of the grid (when visible) corresponding to the conventional
x- and y- axes of the Cartesian plane. By default the grid is
1-by-1. However, by the command \Edges{100}, you can change this
to 100-by-100 and many users find this alternative most
convenient. Place the command \Edges{...} in your style file (or
header) since its effect is is global. Other possible edge values
are 10 and 1000.

  If you use the command \Edges{...} at all, do so with care.  For
if you accidentally delete an \Edges{...} command your labels will
abruptly be badly misplaced and may logically but mysteriously
generate "dimension too big" errors under TeX and "off page" errors
under your driver.  

  You can dictate the edgescale for an individual figure by giving
the scale in brackets immediately after \SetLabels.  Thus, to
import into an article using say \Edge{100} a figure labelled using
another edgescale, say the original 1-by-1 default, you can use
\SetLabels[1]...\endSetLabels.

GETTING IT DOWN PAT

     Complicated labeling deserves the same respect as
complicated mathematics.  Do not expect it to come out perfect the
first time!  What is needed in either case is a mechanism to
repeatedly typeset troublesome pieces.

     One mechanism is always available.  One does complicated
labelling in a separate "test" file involving just the figure being
labelled;  a texpert will know how to \dump TeX's current state as
a temporary format that restarts rapidly at each retry.  Usually,
one then pastes the completed labelled figure back into the main
TeX file, but, of course, one can also \input it as an auxiliary
file.

     If you do not have a TeXpert at handy, here is a first
approximation to an efficient setup. By deletions reduce a copy
of your article to just a few lines before and after the figure.
Now label the figure, and finally, copy and paste the labelled
figure to the original article. Then copy the next figure to label
into this testbed and repeat. The TeXpert can improve the  speed
at which TeX starts up, by compiling a format specifically for
your article; just one caution: best NOT include in the format
ephemeral details of setup like \Set<mydriver>ArtSpecials (from
boxedeps.tex because this reads  figure dimensions which you may
change during your work session.

     An improved mechanism to repeatedly typeset troublesome
pieces is now available on the Macintosh; it is called LinoTeX;
see the same ftp sources.  It could be set up on many types
of computer.

     Before using labelfig.tex to attach labels to a graphics
object inserted using boxedeps.tex or BoxedArt.tex, make it a
firm rule to carefully adjust the bounding box using the trimming
commands of these packages, and also at least tentatively scale
and position the object. Beware of changing the grid inadvertently
after the labels have been positioned.  For example, correcting
the bounding box of a PostScript graphics object can foul up the
labels by changing the coordinate grid to which the labels are
attached. This is particularly true for the trimming  commands of
boxedeps.tex and BoxedArt.tex. However, as noted already, change
of scale is much less disruptive, and modest adjustments should be
well tolerated.

     Sometimes the labels protrude so far from the bounding box
of a figure that the figure has to be repositioned.  Best do this
by ad hoc spacing, say using \hglue and \vglue; altering the
bounding box would create a vicious circle.

     Remember that you are responsible for preventing labels
from overlapping. You are responsible for all label typography
including size and style. A label is really just about anything
that can be put in a TeX box. Note that spaces at the beginning
and end of labels will normally be suppressed; if you really want
them you must protect them with TeX braces.

     This package temporarily sets the \mathsurround parameter
of TeX to zero  while the labels are being affixed. This is done
because nonzero \mathsurround space would influence the position
of left and right aligned labels; then, when a texpert or printer
modifies mathsurround, diagram labeling might be disastrously
altered. There is a small price to pay involving labels that are
formatted as caption boxes including mathematics: you  may want or
need to specify an explicit mathsurround space within the caption
box; it will not influence anything outside.

     Those hostile to the use of * as separator between
the X and Y coordinates of label insertion points, are free to
impose another using \SetXYSeparator{<the new separator>}.  
Americans may prefer "," to "*" since they never use a 
comma as a decimal point; on the other hand, * may be more visible.

APPENDIX (I)  MERGING labelfig.tex LABELS INTO AN EPSF GRAPHICS OBJECT.

     As promised in the introduction, here is a recipe useful for
publishers. It works at least on Macintosh and at least for vectorized
graphics and Adobe type1 fonts.  (There is surely a similar recipe for
PCs under MSWindows.)

 (a)  Use boxedeps.tex utility to integrate the figure given by the eps
file, "x.eps" say, with a visible frame around it.  See
\ShowDisplacementBoxes command in boxedeps.tex.  To get precise results
automatically it is important to use the \Trim... commands of
boxedeps.tex making the "DisplacementBox" neatly fit the figure.

 (b)  Use the TeX printer driver and LaserWriter (versions >= 8.1.1) to
export to an EPSF the DVI page containing the integrated, labelled
figure. You now have an EPS file  "xx.eps"  that contains too much, and at
the wrong scale, and at wrong position.

 (c)  Convert the EPSF to an Adode Illustrator format EPSF using
the shareware utility called epsConvert by Sam Weiss
1993-- (currently $25).

 (d)  In Illustrator (or a compatible program), group the labels and the
"DisplacementBox"; copy them to the clipboard and paste them into "x.ps".
This step requires that all the label fonts be "visible to the Macintosh.

 (e)  Translate and scale the pasted group consisting of the labels plus
the "DisplacementBox" so as to make the "DisplacementBox" the bounding
box of (labelless) figure represented by "x.eps".  At this point the
labels will be correctly placed on the figure "x.eps".

 (f)  Ungroup and delete the "DisplacementBox".  The result is the
desired single EPS file, "x+.eps" say, It contains the original figure
plus its labels.  

     Using grouping and ungrouping appropriately in "x+.eps", a
publisher's staff can very efficiently improve label positions etc.

APPENDIX II)  SOME EXOTIC APPLICATIONS

     The grid of labelfig.tex is analogous to a light-table in
classical page makeup with wax or latex glue.  In principle, you
can use it to compose any page from its indivisible parts.  This
even has some of the artisanal charm of classical paste-up
provided you have a fast screen preview to make the process
"interactive".

     In practice labelfig.tex is a tool for nonstandard jobs.
Here are a few going beyond the labelling already discussed.

(I)  GRAPHICS INTEGRATION.

     This is accomplished by treating the imported graphics
objects as labels.  The underlying graphics object is then
typically an empty  \vbox to <dimension>{\vfill} in a TeX
\midinsert...\endinsert construction.  A label line
might be of the form

   (.1*.1) \\

The exact form of the special command varies from driver to
driver.  However, in the case of encapsulated PostScript graphics
(EPSF norm), by relying on boxedeps.tex, one can have the
following standard syntax (independant of driver  (see
boxedeps.doc for details.
  
  (.1*.1) \BoxedEPSF{MyFigure scaled <scale in mils>}\\

This may be slow since it requires TeX to read the PostScript
file to read bounding box using many complex macros.  So you
may want to try

  (.1*.1) \EPSFSpecial{MyFigure}{<scale in mils>}\\

which is fast and driver independant, but it squashes the
bounding box, normally to its lower left corner.

     Similarly for graphics of the Macintosh PICT norm ---
using BoxedArt.tex (same sources) in place of boxedeps.tex.

     This approach to integration is to be recommended when
one is assembling a composite graphics object.

 (II)  COMMUTATIVE DIAGRAM ENHANCEMENT

     Commutative diagrams or arrays of mathematical objects
connected by arrows of various sorts are common in mathematics.
The mathematical objects require the use of TeX.  Recently TeX
acquired a good collection of arrows of all slopes --- that of
LamSTeX --- plus pwerful macros to build the diagrams.

     However, even the LamSTeX collection is often
inadequate; it lacks for example double shafted arrows, dotted
arrows and curved arrows. Fortunately it is possible to produce
such arrows on an individual basis using sophisticated graphics
programs such as Illustrator and AldusFreehand (both serving
the EPSF norm) or using Metafont (with its public domain norm).
Since the creation of each new arrow is a work of love, you
probably want to limit the number of arrows by using LamSTeX
for most arrows. The 40K commutative diagram module of LamSTeX
has been adapted to work with AmSTeX and a copy may be posted
with LabelFig and related files. Unfortunately no one has yet
offered a version that works with Plain TeX or LaTeX.

       Suffice it here to say that when the exotic arrow has
been somehow imported into TeX, labelfig.tex treats it as a
label that one affixes to the commutative diagram.  Two other
steps will be treated in separate notes, namely the matter of
extracting the dimension specifications for the arrow and the
construction of the arrow --- for these steps are far from
unique and often depend intimately on your computer environment. 
Notes for the Macintosh-Textures-Illustrator combination are
found in the file ExoticArrows.doc.

 (III) NESTING 

Ingenuity pays off in exploiting labelfig.tex. One can
mix graphics and typography quite freely.  labelfig.tex is good
for freeform or overlapping arrangements, while boxedeps.tex (or
BoxedArt.tex) is best for regimented non-overlapping
arrangements --- and the two can be combined.

     The default behavior of labelfig.tex is not ideal 
for nesting objects, because to prevent trouble for beginners
the register for labels is globally cleared when \AffixLabels
concludes.  But there are switches available

      \LabelsGlobal      \LabelsLocal

which change this.  To understand this, extend the above test 
by something like:

 \LabelsLocal
 \SetLabels
    (.5*.5) AAA\\
 \endSetLabels

 {%%% Watch for influence of braces!!
 \SetLabels
    (.5*.5) ZZZ\\
 \endSetLabels
   \AffixLabels{\FirstQuadrant}
 }

   \AffixLabels{\FirstQuadrant}

     There are however potential pitfalls.  Neither
labelfig.tex nor boxedeps.tex has been tested under extreme
conditions. Problems may occur if their procedures are
indiscriminately nested. For boxedeps.tex (not labelfig.tex)
there is a precise cause for worry, namely many of its
variables are "global", which means that TeX braces will not
provide the protection one might expect.

COMMAND SUMMARY FOR labelfig.tex

  Here [...] means optional (one or zero)
       [...]* means any number of such constructs

  \SetLabels
    [[<P>](<X><Sep><Y>) <label> \\]*
  \endSetLabels
  \ShowGrid  % this makes the grid visible (once)
  \AffixLabels{<the figure>}

   --- <P> is tack position, one of eleven or empty
              order irrelevant

                   \L\T      \T      \R\T

                   \L\E      \E      \R\E

                     \L               \R

                   \L\B      \B      \R\B

   --- (<X><Sep><Y>) insertion point;
  <Sep> is separator, = * by default;
  \SetXYSeparator{<Sep>} changes it.
   <X> and <Y> are real numbers

  --- <label> a label to attach 

  --- <the figure> the figure to label 

  \GlobalLabels (default)     
  \LocalLabels  setting for nested constructs.

 \Grids makes ALL grids appear; \HideGrid then makes just next disappear.
 \noGrids returns to default.  The commands are always global.

 \GridLineWidth{<dimension>} adjusts width of grid lines. Default is very
small, to give "hairline" effect. If your grid lines are missing try
setting \GridLineWidth{1pt}.

 \Edges#1 globally changes the edge size of all grids to the numerical 
value #1, which must be 1, 10, 100, or 1000.  The default is 1.

VERSION HISTORY.
 --- Jan 1993: \Edges#1 and [??] option after \SetLabels
 --- July 1992: \Grids, \noGrids, \HideGrid;
       Gridlines become hairlines; \GridLineWidth{<dimension>}.
 --- Oct 1991, Jan 1992: \SetXYSeparator{<Sep>},  \LabelsGlobal,
       \LabelsLocal.
 --- July 1991: first release

Address for bugs and other feedback:

        Raymond S\'eroul
        IREM and Lab. de Typographie Informatise
        Univ. Rene Descartes
        Strasbourg

    Tel 33-88-41-63-45
    Email:  A18645@FRCCSC21.BITNET

        Laurent Siebenmann
        Mathematique, Bat. 425,
        Univ de Paris-Sud,
        91405-Orsay,
        France

    Tel 33-1-6941-7949; 
    Email: lcs@topo.math.u-psud.fr

\usepackage{spconf,amsmath,epsfig,epsf,psfrag,amssymb,amsfonts,latexsym, amsmath,color}
\usepackage{verbatim}
\usepackage[mathscr]{eucal}

\newcommand{\beq}{\begin{equation}}
\newcommand{\eeq}{\end{equation}}

\def\defeq{\stackrel{\Delta}{=}}
\def\Ebb{{\mathbb E}}

\newcommand{\Kmsc}{{\mathscr{K}}}
\newcommand{\Imsc}{{\mathscr{I}}}
\newcommand{\SNR}{\mbox{SNR}}

\definecolor{bgrd}{rgb}{1,1,1}
\definecolor{grey}{rgb}{0.9,0.9,0.6}
\definecolor{gray}{rgb}{0.5,0.5,0.5}

\def\L{{\cal L}}

\title{Optimal Node Density for Two-Dimensional Sensor Arrays}
\name{Youngchul Sung\sthanks{{\scriptsize Y. Sung and H. Yu is
with the Dept. of Electrical Engineering, Korea Advanced Institute
of Science and Technology (KAIST), Daejeon 305-701, South Korea.
Email:ysung@ee.kaist.ac.kr and hjyu@stein.kaist.ac.kr. H. V. Poor
is with the Dept. of Electrical Engineering,  Princeton
University, Princeton, NJ 08544. Email: poor@princeton.edu. The
work of Y. Sung was  supported in part by Brain Korea 21 Project,
the School of Information Technology, KAIST. The work of H. V. Poor was
supported in part by the U. S. National Science Foundation under Grants ANI-03-
38807 and CNS-06-25637.}}, H. Vincent Poor
and Heejung Yu }
\address{}

\begin{document}
\maketitle

\ninept

{\footnotesize
\begin{abstract}
The problem of optimal node density for {\em ad hoc} sensor
networks deployed for making inferences about two
dimensional correlated random fields is considered. Using
a symmetric first order conditional
autoregressive Gauss-Markov random field model, large
deviations results are used to characterize the asymptotic per-node
information gained from the array.  This result then allows an analysis
of the node
density that maximizes the information under an energy constraint,
yielding insights into the trade-offs among the
information, density and energy.

\end{abstract}
}

%{\footnotesize \textbf{\textit{Index Terms-}}  Neyman-Pearson
%detection, error exponent, GMRF}

\section{Introduction} \label{sec:intro}

We consider the design of wireless {\em ad hoc} sensor networks
for making inferences about correlated random fields
 that can model various physical processes, such as
temperature, humidity or the density of a certain gas, in a
two-dimensional (2-D) space. In particular, we consider the
optimal density problem for sensor networks deployed for
statistical inference such as detection or reconstruction of the
underlying field. From the information-theoretic perspective,
statistical inference via sensor networks can be viewed as a problem of
extracting information about an underlying physical process using
networked sensor nodes that consume energy for both sensing and
communication. Thus, the optimal density problem can be formulated
as follows.

\begin{problem} \label{problem:optdensity} Given a sensor network deployed on a
 fixed coverage area of size $2L \times 2L$ and
with total available energy $E$, find the node density $\mu_n$ that
maximizes the total information $I_t$ obtainable from the
network.
\end{problem}

\noindent To address this problem, we model the signal field as a
2-D Gauss-Markov random field (GMRF), and consider the
Kullback-Leibler information (KLI) and mutual information (MI)
\cite{Liese&Vajda:06IT} as ways of quantifying inferential performance. (The
operational meaning of the KLI is given by its appearance as  the error exponent of
the miss probability of Neyman-Pearson detection of the signal
field in sensor noise, whereas that of the MI is given by its role as a measure
of  uncertainty reduction.)  Our approach to determine the total
information obtainable  from a sensor network is based on the large deviations principle
(LDP). That is, for large networks, the total information is
approximately given by the product of the number of sensors and
the asymptotic per-node information, or  the asymptotic information rate.
(The units of these intensive quantities is thus nats/sample.)
Although closed-form expressions for the
asymptotic per-node information are not available for general 2-D
signals, for   the  conditional
autoregression (CAR) model  closed-form
expressions for the asymptotic KLI and MI rates have been determined by the authors in
\cite{Sung&Poor&Yu:08ICASSP}.
  Based on these   expressions for asymptotic information
  rates and their properties, in the current paper we investigate the problem
  of optimal node density.
  It is seen that there exists a density maximizing the total
  information obtainable under an energy constraint.  The optimal density
  is easily obtained numerically, and the behavior of the total information as a function
  of the density is explained.

\vspace{-0.5em}
%%%%%%%%%%%%%%%%%%%%%%%%%%%%%%%%%
\subsection{Related Work}
%%%%%%%%%%%%%%%%%%%%%%%%%%%%%%%%%
\vspace{-0.3em}

The issues of optimal sensor density  and optimal sampling have
been considered  based on LDP in  previous work (e.g.,
\cite{Chamberland&Veeravalli:06IT}). However, most work in this
area is  based on  one-dimensional (1-D) signal or time series models
that do not capture the two-dimensionality of actual spatial
signals. In contrast, our work is based on the LDP results obtained in
\cite{Sung&Poor&Yu:08ICASSP}, where a closed-form expression for
the  asymptotic KLI rate is obtained in the spectral domain. For a
2-D setting, an error exponent was obtained for the detection of
2-D GMRFs in \cite{Anandkumar&Tong&Swami:07ICASSP}, where the
sensors are located randomly and the Markov graph is based on the
nearest neighbor  dependency enabling a loop-free graph. In that
work, however, measurement noise was not considered, unlike the
present analysis.

\vspace{-0.7em}
%%%%%%%%%%%%%%%%%%%%%%%%%%%%%%%%%%%%%%%%%%%%%%%%%%%%%%%%%%%%%%%%%%%%%%%
\section{Signal Model and Background}
\label{sec:systemmodel} \vspace{-0.5em}

\begin{figure}[htbp]
\centerline{
    \begin{psfrags}
    \psfrag{ij}[l]{{\scriptsize $(i,j)$}}
    \psfrag{xij}[c]{{\scriptsize $X_{ij}$}}
    \psfrag{wij}[l]{{\scriptsize $W_{ij}$}}
    \psfrag{yij}[c]{{\scriptsize $Y_{ij}$}}
    \psfrag{Nij}[c]{{\scriptsize Sensor $ij$}}
    \psfrag{r}[c]{{\scriptsize $d_n$}}
    \scalefig{0.33}\epsfbox{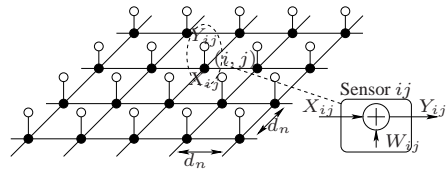}
    \end{psfrags}
} \caption{2-D sensor array on a lattice ${\mathcal I}_n$: Hidden
Markov structure} \label{fig:2dHGMRF}
\end{figure}

In this section, we briefly introduce our previous work
\cite{Sung&Poor&Yu:08ICASSP} relevant to the sensor density
problem. To simplify the problem and gain insight into the 2-D
case, we assume that sensors are located on a 2-D lattice
$\Ic_n=[-n:1:n]^2$, as shown in Fig. \ref{fig:2dHGMRF}, and thus
form a 2-D array. We model the underlying physical process as a
2-D GMRF and assume that each sensor has Gaussian measurement
noise. So, the observation $Y_{ij}$ of Sensor $ij$ on the 2-D
lattice ${\mathcal I}_n$ is given by
\begin{equation} \label{eq:hypothesis2d}
 Y_{ij} = X_{ij}+ W_{ij}, ~~ij \in {\cal I}_n,
\end{equation}
where  $\{W_{ij}\}$ represents independent and identically
distributed (i.i.d.) zero-mean Gaussian measurement noise with
variance $\sigma^2$, and $\{X_{ij}\}$ is a  GMRF on $\Ic_n$,
independent of $\{W_{ij}\}$. Note that the observation samples
form a 2-D hidden GMRF on $\Ic_n$. In the following, we summarize
our relevant LDP results on GMRFs that will be useful in the
sequel.

\begin{definition}[GMRF \cite{Rue&Held:book}]\label{def:GMRF}
A random vector $\Xbf=(X_1,X_2,\cdots,$ $X_n)$ $\in {\mathbb R}^n$
is a Gauss-Markov random field with respect to (w.r.t.). a labelled graph ${\mathcal
G}=({\mathcal \nu},{\mathcal E})$ with mean vector $\mubf$ and
precision matrix $\Qbf
>0$, if its probability density function is given by
{\footnotesize
\begin{equation}
p(\Xbf) = (2\pi)^{-n/2}|\Qbf|^{1/2}\exp\left( - \frac{1}{2}
(\Xbf-\mubf)^T \Qbf (\Xbf-\mubf) \right),
\end{equation}}
and $Q_{lm} \ne 0 \Longleftrightarrow \{l,m\} \in {\mathcal
E}~\mbox{for all}~ l \ne m$.  Here, ${\mathcal \nu}$ is  the set
of all nodes $\{1,2,\cdots, n\}$ and ${\mathcal E}$ is the set of
edges connecting pairs of nodes, which represent the conditional
dependence structure.
\end{definition}

\noindent Note that the 2-D indexing scheme $ij$ in
(\ref{eq:hypothesis2d}) can be properly converted to an 1-D scheme
to apply Definition \ref{def:GMRF}. From here on, we use the 2-D
indexing scheme for convenience.

\begin{definition}[The Conditional Autoregression (CAR)]
A GMRF $\{X_{ij}\}$ is said to be a conditional autoregression if
it is specified using a set of full conditional normal
distributions with means and precisions: {\footnotesize
\begin{eqnarray}
\Ebb \{ X_{ij}|\Xbf_{-ij}\} &=&  -\frac{1}{\theta_{00}}
\sum_{i^\prime j^\prime \in {\mathcal I}_\infty \ne 00}
\theta_{i^\prime j^\prime} X_{i+i^\prime,j+j^\prime}, \label{eq:condMean2DInf}\\
\mbox{Prec}\{X_{ij}|\Xbf_{-ij}\} &=& \theta_{00} > 0,
\label{eq:condPrec2DInf}
\end{eqnarray}}
where $\Xbf_{-ij}$ denotes the set of all variables except
$X_{ij}$.
\end{definition}

\noindent By imposing first order symmetry on the correlation
structure, we have the symmetric first order conditional
autoregression (SFCAR) defined by the conditions {\small
\begin{eqnarray*}
\Ebb \{ X_{ij}|\Xbf_{-ij}\} &=&  \frac{\lambda}{\kappa} (X_{i+1,j}+X_{i-1,j}+X_{i,j+1}+X_{i,j-1}),\\
\mbox{Prec}\{X_{ij}|\Xbf_{-ij}\} &=& \kappa > 0,
\end{eqnarray*}}
where $0 \le \lambda \le \frac{\kappa}{4}$. Here,
$\theta_{00}=\kappa$ and $\theta_{1,0} = \theta_{-1,0} =
\theta_{0,1} = \theta_{0,-1} = -\lambda$.  The SFCAR model is the
2-D extension of the 1-D autoregressive (AR) model that is widely
used to model basic correlation in 1-D. Here in the 2-D case we have
symmetric conditional dependence on four neighboring nodes in the
four (planar) directions, capturing basic 2-D correlation structure. It
can be shown that the GMRF defined by the SFCAR model is a zero-mean
stationary Gaussian process on ${\mathcal I}_\infty$ with
power spectral density \cite{Rue&Held:book}
\begin{equation}
f(\omega_1,\omega_2) = \frac{1}{4\pi^2 \kappa (1 - 2 \zeta
\cos\omega_1 - 2 \zeta \cos\omega_2)},
\end{equation}
where the {\em edge dependence factor} $\zeta$ is defined as
\vspace{-0.5em}
\begin{equation}
\zeta
\defeq \frac{\lambda}{\kappa}, ~~~~ 0 \le \zeta \le 1/4.
\end{equation}
The SFCAR model is useful especially because the correlation
strength is captured in this single quantity $\zeta$ for SFCAR
signals, which enables us to investigate the per-node information
as a function of the field correlation. Here, $\zeta =0$ corresponds
to the i.i.d. case, whereas $\zeta =1/4$ corresponds to the
perfectly correlated case. Henceforth, we assume that the 2-D
stochastic signal $\{X_{ij}\}$ in (\ref{eq:hypothesis2d}) is given
by a stationary GMRF defined by the SFCAR model, as $n\rightarrow
\infty$.  The signal power $P\defeq \Ebb\{X_{00}\}^2$
($=\Ebb\{X_{ij}^2\} ~\forall~i,j$) is obtained using the inverse
Fourier transform, and is given by $P
 = \frac{2K(4\zeta)}{\pi \kappa}, ~\left(0 \le \zeta \le
\frac{1}{4} \right)$, where $K(\cdot)$ is the complete elliptic
integral of the first kind \cite{Besag:81JRSS}. Thus, the
measurement SNR is given by $\mbox{SNR} = \frac{P}{\sigma^2} =
\frac{2K(4\zeta)}{\pi \kappa \sigma^2}$.

\subsection{Large System Analysis: Per-Node Information}

 The key idea behind the large system analysis here is that, under the stationarity assumption,
  the amounts of information from the
 node become identical regardless of sensor location as the
 network size grows, and  the total amount of
information is given approximately by the product of the number of
sensor nodes and the (asymptotic) per-node information.  The
asymptotic per-node KLI and per-node MI are defined as
\[
\Kmsc_s = \lim_{n \rightarrow \infty} \frac{1}{|\Ic_n|} \log
\frac{p_{0}}{p_{1}}(\{Y_{ij}, ij \in \Ic_n\}) ~\mbox{a.s.
under}~p_{0}, ~~\mbox{and}
\]
\[
\Imsc_s = \lim_{n \rightarrow \infty} \frac{1}{|\Ic_n|}
I(\{X_{ij}, ij \in \Ic_n\};\{Y_{ij}, ij\in \Ic_n\}),~~~~~~~~~
\]
respectively. For the MI, the signal model (\ref{eq:hypothesis2d})
is  applicable directly, whereas for the KLI the probability
density functions of the null (noise-only) and alternative
(signal-plus-noise) distributions are those given under the respective
models
\begin{eqnarray}
p_0 (Y_{ij}) &:&  Y_{ij} = W_{ij} , ~~ij \in {\cal I}_n, \nonumber\\
p_1(Y_{ij})  &:& Y_{ij} = X_{ij}+ W_{ij}, ~~ij \in {\cal I}_n.
\label{eq:KLIp0p1}
\end{eqnarray}
The following closed-form expressions for the asymptotic per-node
information in the spectral domain have been obtained in
\cite{Sung&Poor&Yu:08ICASSP} by exploiting the spectral structure
of the CAR signal and the relationship between the eigenvalues of
block circulant and block Toeplitz matrices representing 2-D
correlation structure.

\begin{theorem}\label{theo:eeSFA}
Under the 2-D SFCAR signal model, the asymptotic per-node KLI
$\Kmsc_s$ and per-node MI $\Imsc_s$  are given by {\tiny
\begin{eqnarray}
\Kmsc_s &=& \frac{1}{4\pi^2} \int_{-\pi}^{\pi} \int_{-\pi}^{\pi}
\biggl( \frac{1}{2}\log \left(1+\frac{ \mbox{SNR}}{
(2/\pi)K(4\zeta) (1 - 2 \zeta \cos\omega_1 - 2 \zeta
\cos\omega_2)}\right) \nonumber\\
&& ~~~~~~+\frac{1}{2} \frac{1}{1+\frac{ \mbox{SNR}}{
(2/\pi)K(4\zeta) (1 - 2 \zeta \cos\omega_1 - 2 \zeta
\cos\omega_2)}} -\frac{1}{2} \biggl)d\omega_1d\omega_2.
\label{eq:aKLIR_SFA}
\end{eqnarray}
} and {\tiny
\begin{equation}
\Imsc_s = \frac{1}{4\pi^2} \int_{-\pi}^{\pi} \int_{-\pi}^{\pi}
\frac{1}{2}\log \left(1+\frac{ \mbox{SNR}}{ (2/\pi)K(4\zeta) (1 -
2 \zeta \cos\omega_1 - 2 \zeta \cos\omega_2)}\right)
d\omega_1d\omega_2, \label{eq:aMLIR_SFA}
\end{equation}
} respectively.
\end{theorem}

\begin{figure}[htbp]
\centerline{ \SetLabels
\L(0.25*-0.1) (a) \\
\L(0.72*-0.1) (b) \\
\endSetLabels
\leavevmode
%\ShowGrid
\strut\AffixLabels{ \scalefig{0.235}\epsfbox{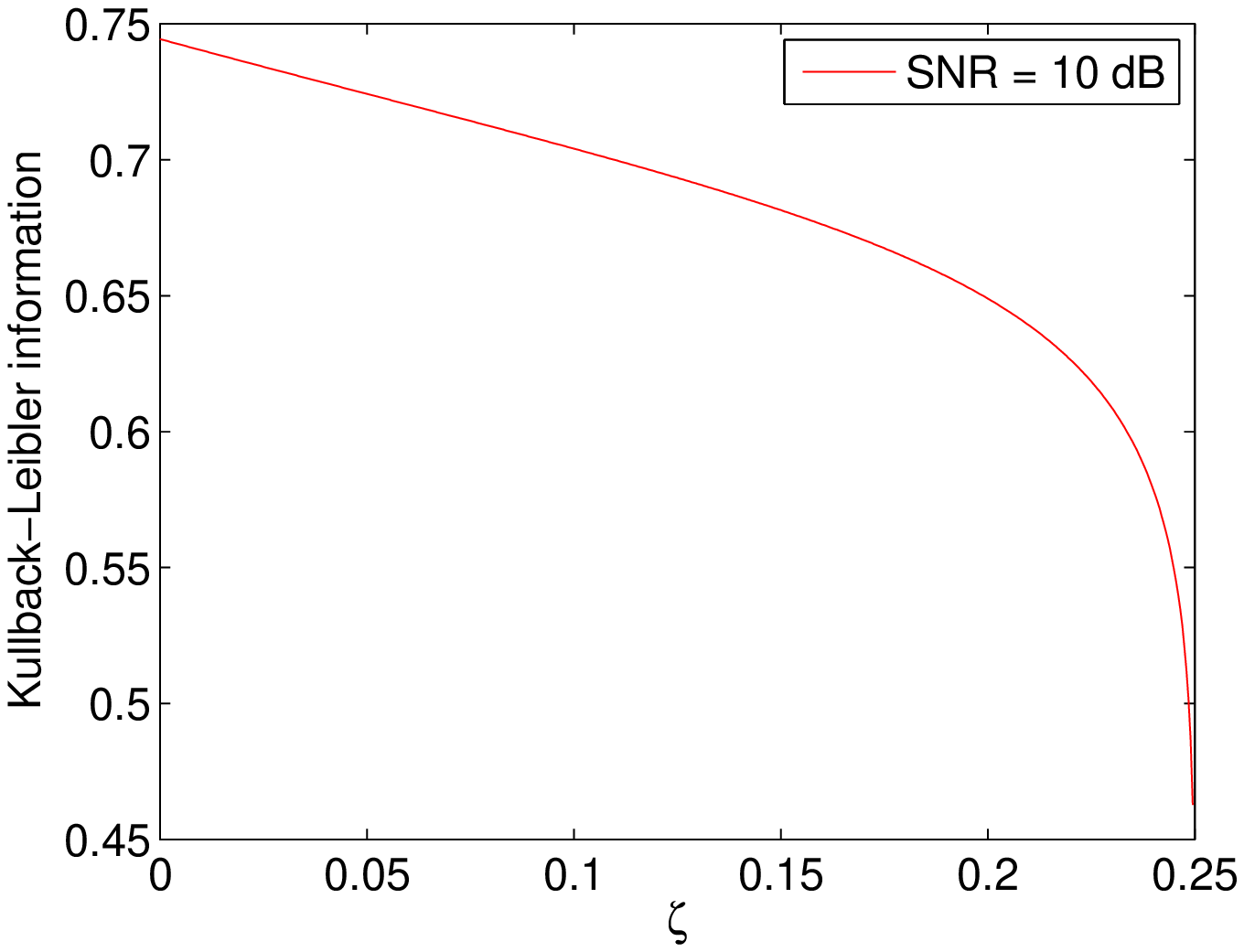}
\scalefig{0.235}\epsfbox{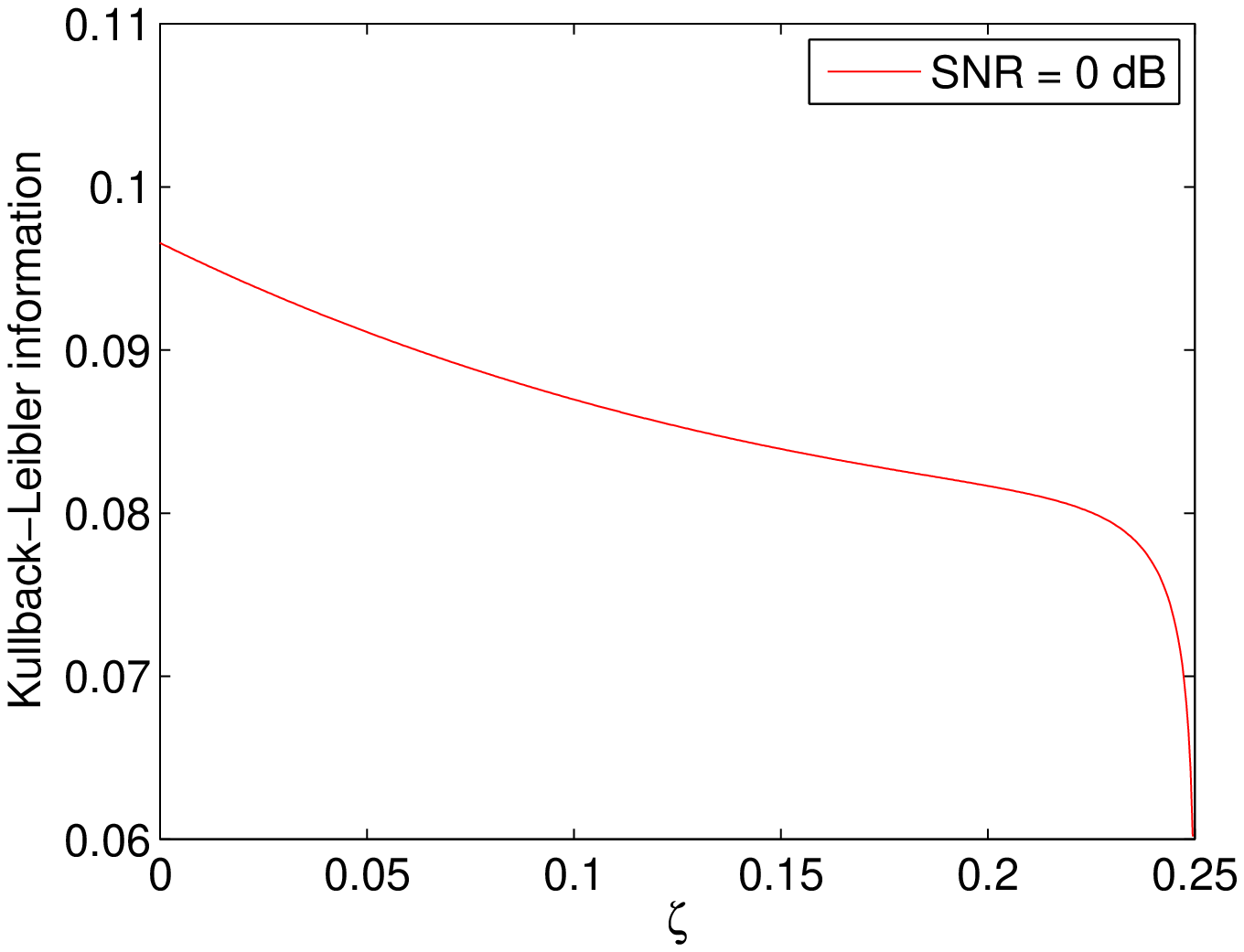} } } \vspace{0.5cm}
\centerline{ \SetLabels
\L(0.25*-0.1) (c) \\
\L(0.72*-0.1) (d) \\
\endSetLabels
\leavevmode
%\ShowGrid
\strut\AffixLabels{ \scalefig{0.235}\epsfbox{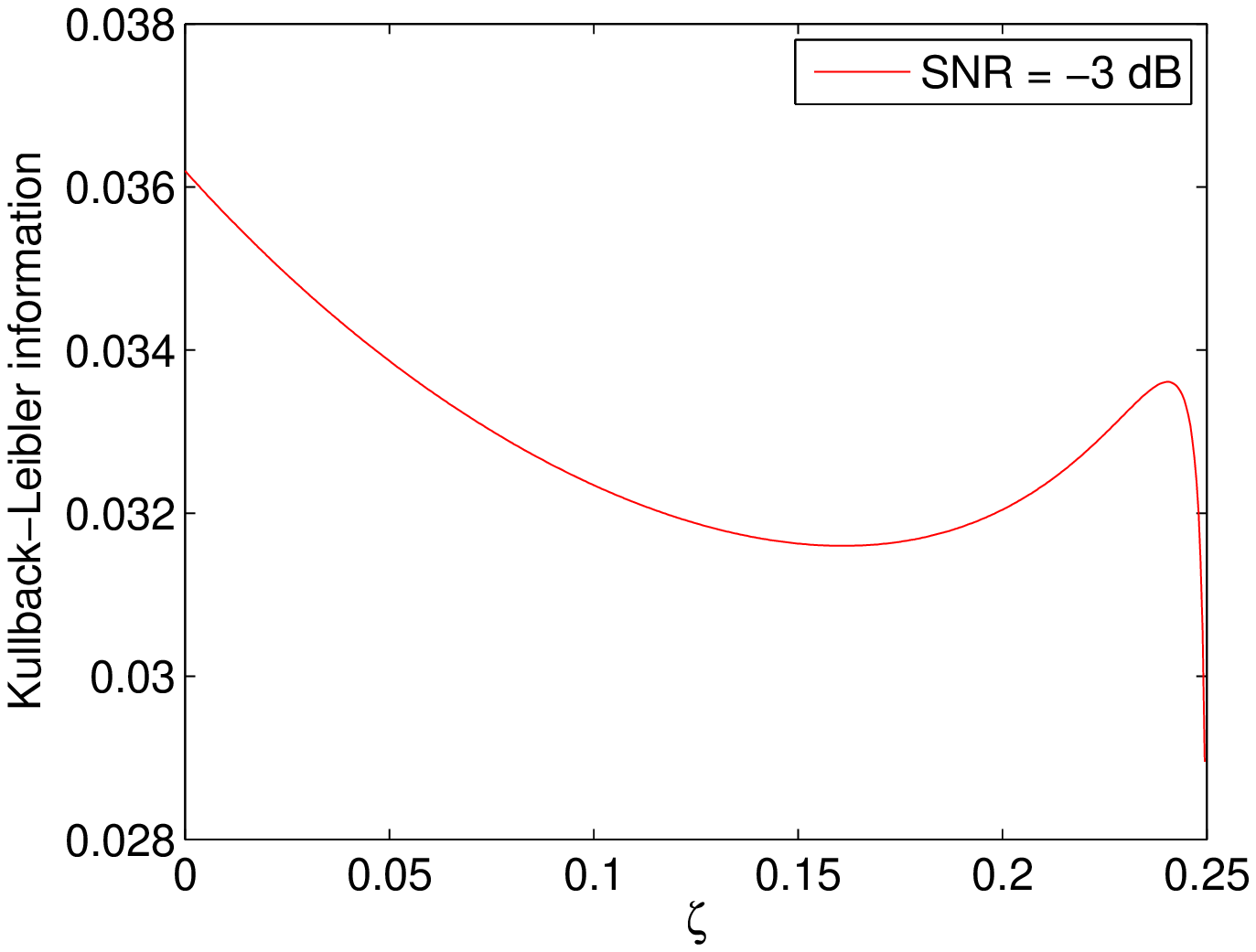}
\scalefig{0.235}\epsfbox{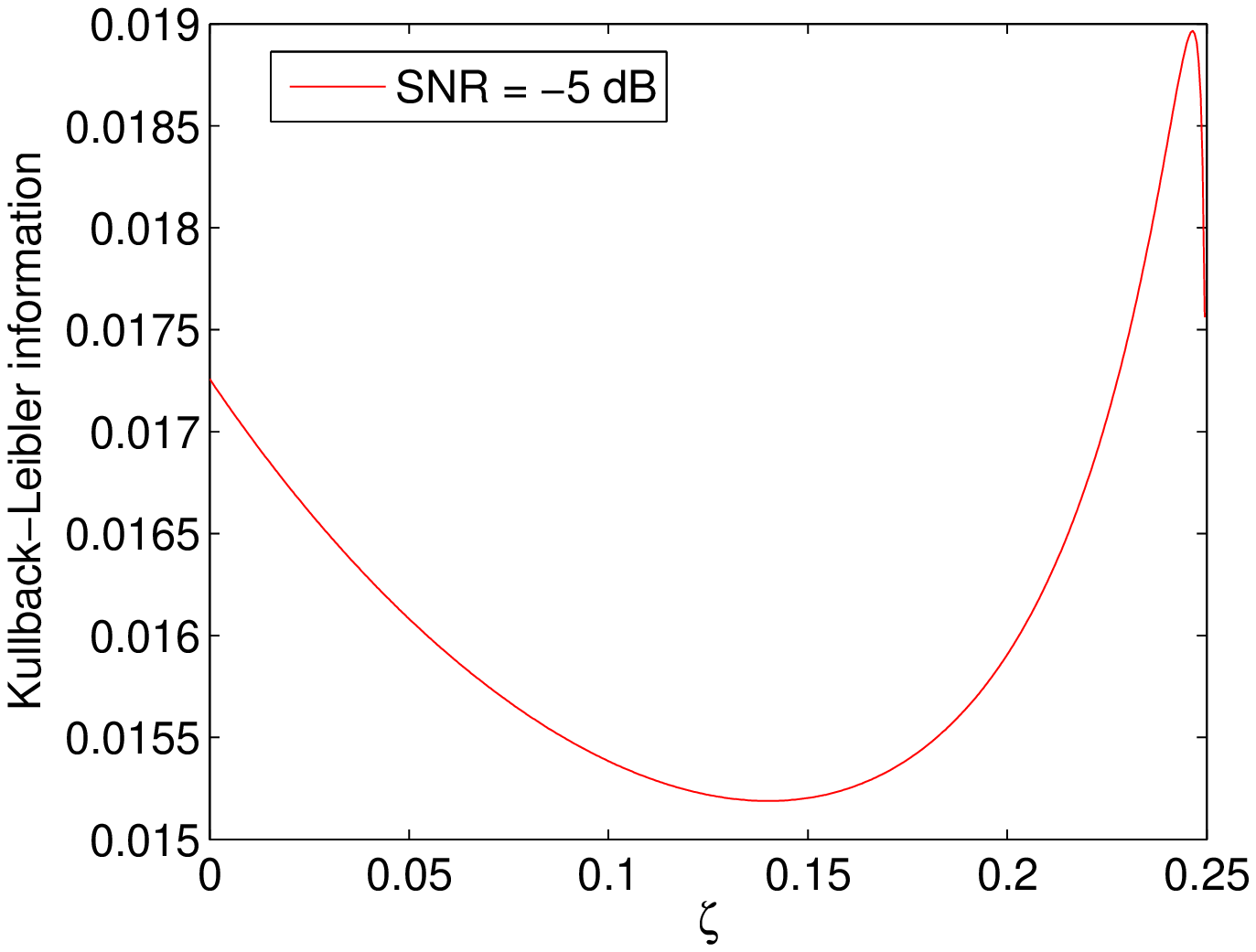} } } \vspace{0.5cm}
\caption{$\Kmsc_s$ as a function of $\zeta$: (a) SNR = 10 dB, (b)
SNR = 0 dB, (c) SNR = -3 dB, (d) SNR = -5 dB (from
\cite{Sung&Poor&Yu:08ICASSP})}
 \label{fig:KcsVsZeta}
\end{figure}

\noindent Note that the SNR and correlation are separated in
(\ref{eq:aKLIR_SFA})-(\ref{eq:aMLIR_SFA}), which enables us to
investigate the effects of each term separately. With regard to
$\Kmsc_s$ and $\Imsc_s$ as functions of $\zeta$, it is readily
seen from Theorem \ref{theo:eeSFA} that $\Kmsc_s$ and $\Imsc_s$
are continuously differentiable $C^1$ functions of the edge
dependence factor $\zeta$ ($0 \le \zeta \le 1/4$) for a given SNR
since $f:x \rightarrow K(x)$ is a continuously differentiable
$C^\infty$ function for $0 \le x < 1$ \cite{Erdelyi:53book}. Fig.
\ref{fig:KcsVsZeta} shows $\Kmsc_s$ as a function of $\zeta$ for
several different SNR values.
It is seen in the figure that at high SNR $\Kmsc_s$ decreases
monotonically as the correlation becomes strong, i.e., $\zeta
\rightarrow 1/4$. At low SNR, on the other hand, correlation is
beneficial to the performance. $\Imsc_s$ shows similar behaviors
even if it is not shown here.

 With regard to $\Kmsc_s$  and $I_s$ as functions of SNR, the
 behavior  is given by the following theorem from
 \cite{Sung&Poor&Yu:08ICASSP}.

\begin{theorem} \label{theo:KLIsvsSNR}
$\Kmsc_s$  and $\Imsc_s$ are continuous and monotonically
increasing as SNR increases for a given edge dependence factor $0
\le \zeta < 1/4$. Moreover, $\Kmsc_s$ and $\Imsc_s$ increase
linearly with respect to $\frac{1}{2}\log \mbox{SNR}$ as $\SNR
\rightarrow \infty$. As SNR decreases to zero, on the other hand,
$\Kmsc_s$ and $\Imsc_s$ decrease to zero with convergence rates
\begin{eqnarray}
\Kmsc_s(\SNR) &=& c\cdot \SNR^2 + o(\SNR^2),  \label{eq:SNR0K}\\
\Imsc_s(\SNR) &=& c^\prime \cdot \SNR + o(\SNR), \label{eq:SNR0I}
\end{eqnarray}
respectively, for some constants $c$ and $c^\prime$.
\end{theorem}

\section{Ad Hoc Sensor Networks: Optimal Density}

Based on the  results in the previous sections, we now address the
optimal density problem given in Section \ref{sec:intro}.

\subsection{Physical correlation model}
\label{subsec:physicalmodel}

As we vary the node density for a given area with size $2L\times
2L$, the sensor spacing $d_n$ changes. In turn, the edge
dependence factor between two adjacent samples varies for given
physical diffusion parameters.  So, we first derive the
relationship between sensor spacing $d_n$ and the edge dependence
factor $\zeta$ for the SFCAR. The physical correlation for the
SFCAR model is obtained by solving the continuous-index equivalent
given by the 2-D stochastic Laplace equation
\cite{Whittle:54Biometrika} {\footnotesize
\begin{equation} \label{eq:laplaceSDE}
\left[ \left( \frac{\partial}{\partial x}\right)^2 +\left(
\frac{\partial}{\partial y} \right)^2 - \alpha^2 \right]X(x,y) =
u(x,y),
\end{equation}} where $u(x,y)$ is the 2-D white zero-mean
Gaussian perturbation and $\alpha > 0$ is the diffusion rate. By
solving this equation, the edge correlation factor $\rho$ is given,
as a function of the sensor spacing $d_n$, by
\cite{Whittle:54Biometrika}
\begin{equation} \label{eq:2DcorrelationFunc}
\rho \defeq \frac{\Ebb\{ X_{00}X_{10}
 \}}{\Ebb\{X_{00}^2  \}}= g(d_n) = \alpha
d_n K_1(\alpha d_n),
\end{equation}
where $K_1(\cdot)$ is the modified Bessel function of the second
kind. The correlation function (\ref{eq:2DcorrelationFunc}) can be
regarded as the representative  correlation in 2-D, similar to the
exponential correlation function $e^{-Ad_n}$ in 1-D. Both
functions decrease monotonically w.r.t. $d_n$. However, the 2-D
correlation function is flat at $d_n=0$
\cite{Whittle:54Biometrika}. Further, we have a mapping $g:\rho
\rightarrow \zeta$ from the edge correlation factor $\rho$ to the
edge dependence factor $\zeta$, given by
\cite{Sung&Poor&Yu:08ITsub}
\begin{equation} \label{eq:ZetaVsRho}
\rho =  \frac{(2/\pi)K(4\zeta)-1}{4 (2/\pi)\zeta K(4\zeta)} =:
h^{-1}(\zeta),
\end{equation}
which maps zero and one to zero and 1/4, respectively. Combining
(\ref{eq:2DcorrelationFunc}) and (\ref{eq:ZetaVsRho}), we have a
mapping $\zeta = h(g (d_n))$ from the sensor spacing $d_n$ to
$\zeta$ for the SFCAR model.

%\vspace{-1em}
\subsection{Density Analysis}
%\vspace{-0.5em}

The assumptions for the planar {\em ad hoc} sensor network that we
consider is summarized in the following.
\begin{itemize}
\item[(A.1)] $(2n+1)^2$ sensors are located on the lattice $\Ic_n
= [-n:1:n]^2$ with spacing $d_n$, as shown in Fig.
\ref{fig:2dHGMRF}, and a fusion center is located at the center
$(0,0)$. The observation samples $\{Y_{ij}\}$ at sensors form a
2-D hidden SFCAR GMRF on the lattice, and the correlation
functions are given by (\ref{eq:2DcorrelationFunc}) - (\ref{eq:ZetaVsRho}).  %

\item[(A.2)] The fusion center collects the measurement from all
nodes using minimum hop routing. A hop count of $|i|+|j|$ is
required for minimum hop routing to deliver $Y_{ij}$ to the fusion
center.

\item[(A.3)] The communication  energy per edge is given by
$E_{c}(d_n) = E_0 d_n^{\nu}$, where  $\nu \ge 2$ is the
 attenuation  factor of wireless propagation in the physical layer.

\item[(A.4)] Sensing requires energy, and the sensing energy per
node is denoted by $E_{s}$.  Further, we assume that the {\em
measurement} SNR  increases linearly w.r.t. $E_{s}$, i.e., $\SNR =
\beta E_s$  for some constant $\beta$.\footnote{
 Suppose that $E_1$ joules are
required for one sensing to obtain one sample $Y_{ij}(m) =
X_{ij}(m) + W_{ij}(m)$ at sensor $ij$ and the measurement SNR of
this sample is $\SNR_1$. Now assume that we obtain $M$ samples
($m=1,\cdots,M$) using $M$ subsensors at the same location $ij$
simultaneously, requiring $M\cdot E_1$ joules, and we take an
average  of these $M$ samples at sensor $ij$, yielding an
effective sample $Y_{ij}=(1/M)\sum_m Y_{ij}(m)$ of SNR of
$M\SNR_1$ assuming that the measurement noise is i.i.d. across the
subsensors.}
\end{itemize}

\noindent The density optimization under the energy constraint can
be solved using our large system analysis in the previous sections
 assuming the asymptotic result is still valid in low density
case. The total amount $I_t$ of information is given by
{\footnotesize
\begin{equation} \label{eq:adhocTotalInfo}
I_t = (2n+1)^2 \Kmsc_s(\mbox{SNR},d_n) ~~\mbox{or}~~ I_t =
(2n+1)^2 \Imsc_s(\mbox{SNR},d_n),
\end{equation}}
for KLI or MI, respectively.  The total energy $E$ required for
data collection  is given by {\footnotesize
\begin{eqnarray}
E &=& (2n+1)^2 E_s + E_c(d_n) \sum_{i=-n}^n\sum_{j=-n}^n
(|i|+|j|),\nonumber\\
&=& (2n+1)^2 E_s + 2n(n+1)(2n+1) E_{c}(d_n).
\label{eq:adhocTotalEnergy}
\end{eqnarray}}
Thus,  Problem 1 can be reformulated as {\scriptsize
\begin{eqnarray}
\mu_{n}^* &=& \mathop{\arg \max}_{\mu_n} ~(2L)^2 \mu_n \Kmsc_s
(\mbox{SNR}(E,\mu_n), d_n(\mu_n)),\label{eq:optimaldensityconstraint}\\
&& \mbox{s.t.} ~ (2n+1)^2 E_s (\mu_n) + 2n(n+1)(2n+1)
E_c(d_n(\mu_n)) \le E, \nonumber
\end{eqnarray}}
where the sensing energy $E_s$ as well as $n$ and $d_n$ are
functions of the node density $\mu_n$.  From $\mu_n ~(=
(2n+1)^2/(2L)^2)$, we first calculate $n$ and then $d_n = L/n$.
When $d_n$ is determined, $E_c(d_n)$ is obtained from the
propagation parameters $E_0$ and $\nu$, and then $E_s(\mu_n)$ is
obtained from the constraint in
(\ref{eq:optimaldensityconstraint}). Once $E_s(\mu_n)$ is
determined, the measurement SNR is calculated using Assumption
{\em (A.5)}, i.e., SNR = $\beta E_s$ and finally we evaluate the
per-sensor information $\Kmsc_s(\SNR, \zeta(\rho(d_n)))$ and
$\Imsc_s(\SNR, \zeta(\rho(d_n)))$ from Theorem \ref{theo:eeSFA}.

\begin{figure}[htbp]
\centerline{ \SetLabels
\L(0.24*-0.1) (a) \\
\L(0.74*-0.1) (b) \\
\endSetLabels
\leavevmode
%\ShowGrid
\strut\AffixLabels{ \scalefig{0.25}\epsfbox{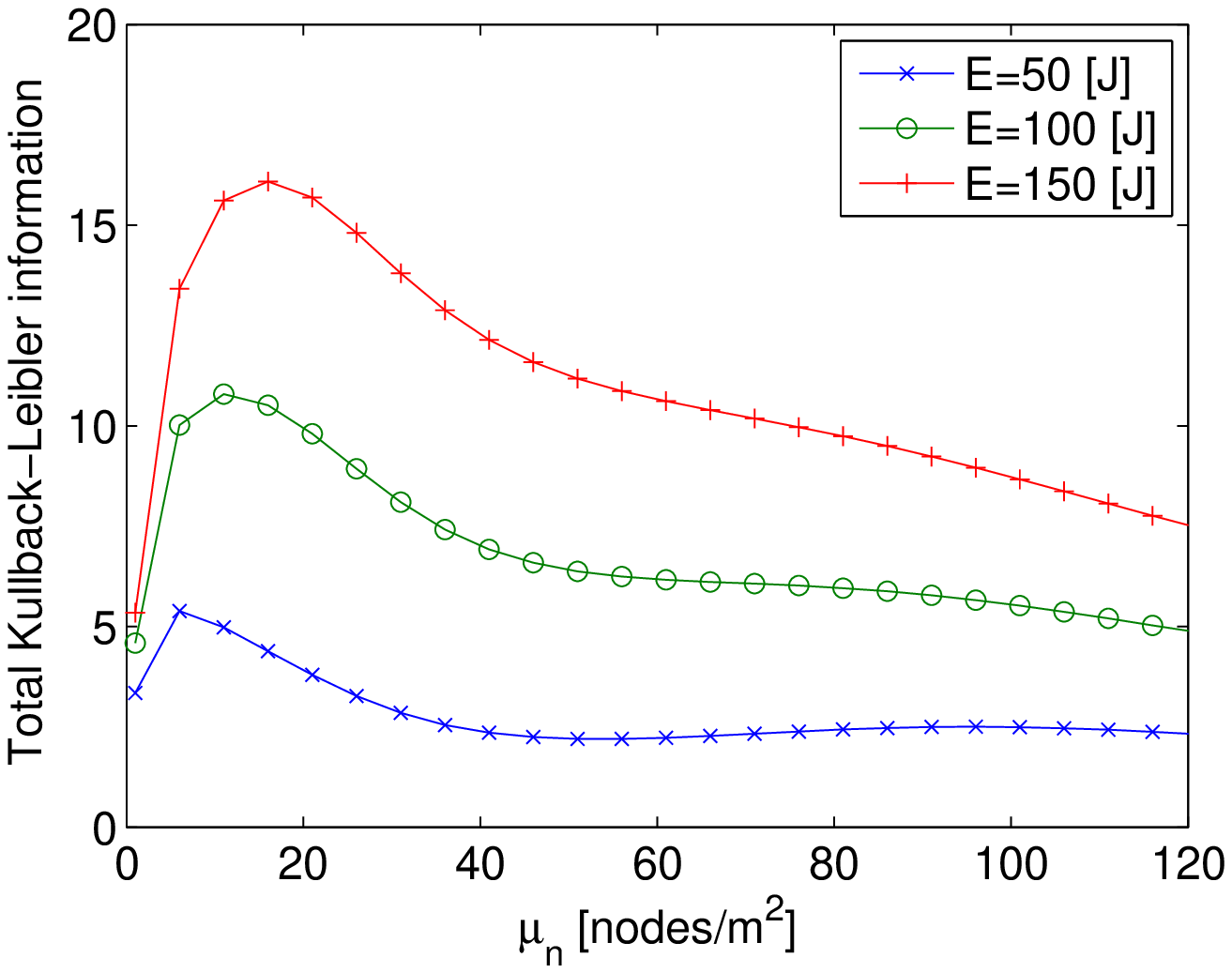}
\scalefig{0.25}\epsfbox{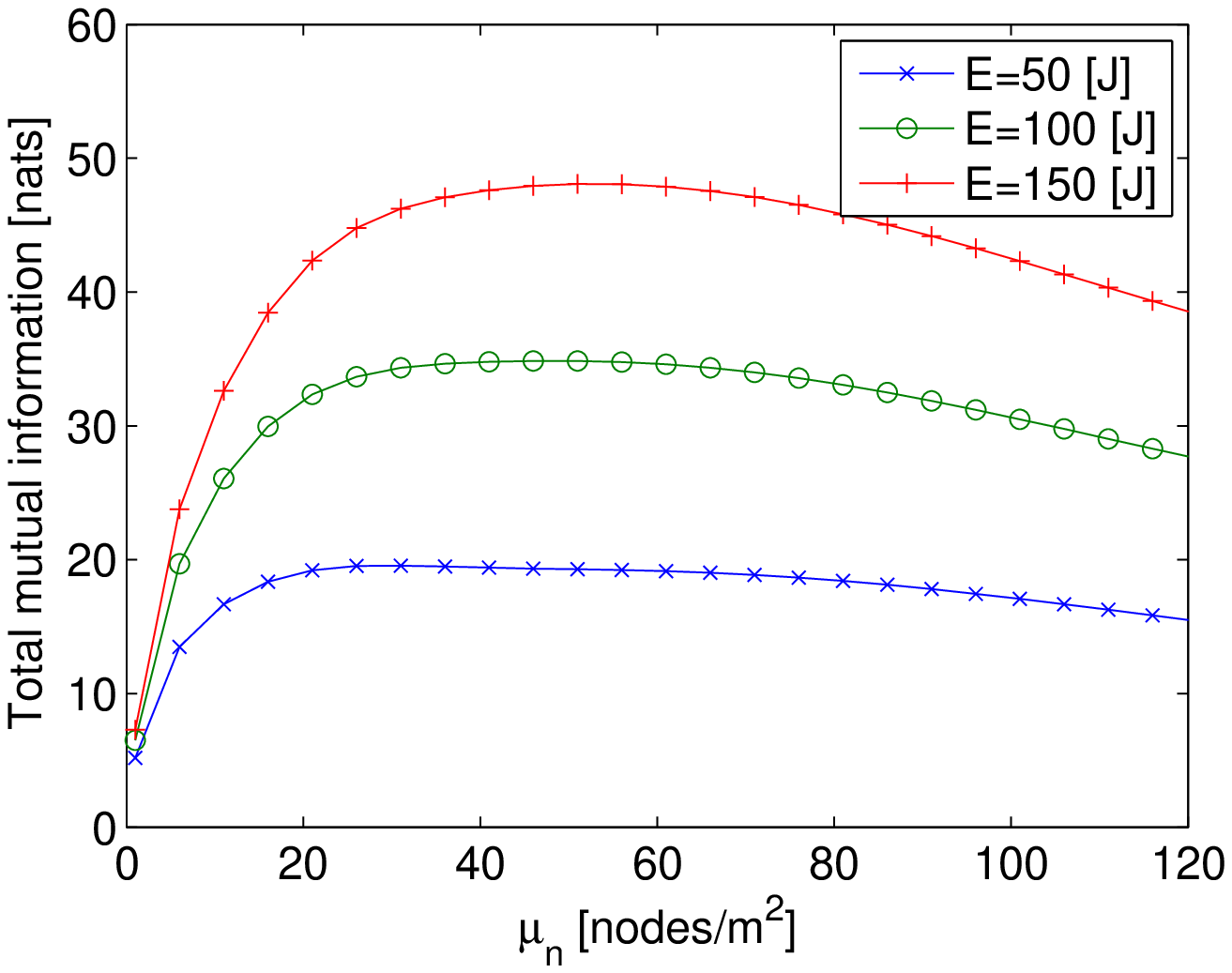} } } \vspace{0.3cm}
 \caption{(a) total KLI vs. density  and (b) total MI vs. density}
\label{fig:InformationVsDensity}
\end{figure}

Fig. \ref{fig:InformationVsDensity} shows the total information
obtainable from 2 $\times$ 2 square meter area as we vary the node
density $\mu_n$ with a fixed total energy budget of  $E$ joules. Other
parameters that we use are given by
\[
\alpha = 100 ~(\mbox{diffusion rate}), ~\beta=1, ~ E_0 = 0.1~
\mbox{and}~ \nu = 2.
\]
Here, the values of $E$, $E_0$ and $\beta$ are chosen so that the
minimum and maximum per-sensor sensing SNR's are roughly -10 to 10
dB for maximum and minimum densities, respectively.  The diffusion
rate $\alpha=100$ is selected for the edge correlation coefficient
$\rho$ to vary from almost zero to 0.6 as the node density
changes. It is seen in the figure that there is an optimal density
for each value of $E$ for both information measures.  It is also
seen that the total KLI is sensitive to the density change whereas
the total MI is less sensitive. The existence of the optimal
density is explained as follows. At low density, we have only a
few sensors in the area. So, the energy for communication is not
large due to the small number of communicating nodes  and most of
the energy is allocated to the sensing energy; the per-node
sensing energy is even higher due to the small number of sensors.
However, the per-node information increases only logarithmically
w.r.t. the sensing energy or SNR by Theorem \ref{theo:KLIsvsSNR},
and this logarithmic gain cannot compensate for the loss in the
number of sensors. Therefore, low density yields very poor
performance, and large gain is obtained initially as we increase
the density from very low values as seen in Fig.
\ref{fig:InformationVsDensity}. As we further increase the
density, on the other hand, and the per-node sensing energy or SNR
decreases due to the increase in the overall communication and the
increase in the number of sensor nodes, and the measurement SNR is
eventually at low SNR regime, where (\ref{eq:SNR0K}) and
(\ref{eq:SNR0I}) hold. From (\ref{eq:adhocTotalEnergy}), we have
\begin{equation}
E_s (\mu_n) =\beta^{-1}\SNR = O(n^{-2})
\end{equation}
for fixed $E$ and $E_c =E_0 (L/n)^2$, as $n \rightarrow \infty$.
By the low SNR behavior of $\Kmsc_s$ given by (\ref{eq:SNR0K}),
the behavior of the total Kullback-Leibler information is given by
\[
\mbox{Total KLI}= (2L)^2 \mu_n \Kmsc_s = O(n^2 n^{-4}) = O(n^{-2})
= O(\mu_n^{-1})
\]
 and by (\ref{eq:SNR0I})  the total mutual information is
given by
\[
\mbox{Total MI} = (2L)^2 \mu_n \Imsc_s = O(n^2 n^{-2}) = O(1).
\]
This explains the initial decay after the peak in Fig.
\ref{fig:InformationVsDensity} (a) and quite flat curve in Fig.
\ref{fig:InformationVsDensity} (b).  In the above equations,
however, the effect of $\zeta$  on $\Kmsc_s$ and $\Imsc_s$ is not
considered. As the node density increases, the sensor spacing
decreases and the edge dependence factor $\zeta$ increases for a
given diffusion rate $\alpha$. The behavior of the per-node
information as a function of $\zeta$ is shown in Fig.
\ref{fig:KcsVsZeta}. Note in Fig. \ref{fig:KcsVsZeta} that the
per-node information  has a second lobe at strong correlation at
low SNR while at high SNR it  decreases monotonically as the
correlation becomes strong. The benefit of sample correlation is
evident in the low energy case ($E=50 [\mbox{J}]$) in
\ref{fig:InformationVsDensity} (a); the second peak  around $\mu_n
= 95$ [nodes/$m^2$] is observed. Note that the second peak is not
so significant. Since the per-node  information decays to zero as
$\zeta \rightarrow 1/4$ eventually, the total amount of
information decreases eventually, as seen in the right corner of
the figure, as we increases the node density.

\section{Conclusions}
\label{sec:conclusion}

We have considered the design of 2-D arrays of networked sensors
for making   inferences about 2-D correlated random fields.
Under the SFCAR GMRF model, the  density maximizing the
total information obtainable from the network under an energy
constraint has been investigated. We have seen that  such an optimal
 density exists. At
low density, the amount of  information gathered is small because
the logarithmic increase in the per-node information w.r.t. energy
cannot compensate for the loss in the number of sensor nodes. At
high density, on the other hand, the performance degrades mainly
due to too much correlation between samples and low sensing
energy. The optimal node density effects a trade-off between these
two effects.

%%%%%%%%%% References %%%%%%%%%%%%%%%%%%%%%%%%%%%%%%%%%%%%%%%%%%%%%%%%%%
%{%\scriptsize
%\bibliographystyle{ieeetr}
%\bibliography{referenceBibs} %{IEEEabrv,referenceBibs}
%}

{%\scriptsize
\bibliographystyle{ieeetr}

}

\end{document}